\documentclass[twocolumn,showpacs,preprintnumbers,amsmath,amssymb]{revtex4}
\usepackage{amssymb}
\usepackage{graphicx}
\usepackage{dcolumn}
\usepackage{bm}
\usepackage{amsmath}
\usepackage{amsfonts}%
\providecommand{\U}[1]{\protect\rule{.1in}{.1in}}
\begin{document}

\title{Nonlinear preferential rewiring in fixed-size networks as a diffusion process}
\author{Samuel Johnson, Joaqu\'{i}n J. Torres, and Joaqu\'{i}n Marro\\{\small Departmento de Electromagnetismo y F\'{\i}sica de la Materia,}\\{\small and Instituto Carlos I de F\'{\i}sica Te\'{o}rica y Computacional,}\\{\small University of Granada, Spain.}}

\begin{abstract}
We present an evolving network model in which the total numbers of nodes and
edges are conserved, but in which edges are continuously rewired according to
nonlinear preferential detachment and reattachment. Assuming power-law kernels
with exponents $\alpha$ and $\beta$, the stationary states the degree
distributions evolve towards exhibit a second order phase transition -- from
relatively homogeneous to highly heterogeneous (with the emergence of starlike
structures) at $\alpha=\beta$. Temporal evolution of the distribution in this
critical regime is shown to follow a nonlinear diffusion equation, arriving at
either pure or mixed power-laws, of exponents $-\alpha$ and $1-\alpha$.

\end{abstract}



\maketitle

Complex systems may often be described as a set of nodes with edges connecting some of them
-- the \textit{neighbours --} (see, for instance, Refs.\cite{gen,gen1,gen2}).
The number of edges a particular node has is called its degree, $k$. The study of such large 
networks is usually made simpler by considering statistical properties, e.g.,
the degree distribution, $p(k)$ (probability of finding a node with a
particular degree). It turns out that a high proportion of real-world networks
follow power-law degree distributions, $p(k)\sim k^{-\gamma}$ -- referred to
as \textit{scale-free} due to their lack of a characteristic size. Also, many
of them have their edges placed among the nodes apparently in a random way
-- i.e., there is no correlation between the degree of a node and any other of
its properties, such as the degrees of its neighbours. Barab\'{a}si and Albert \cite{Barabasi} applied the mechanism of \textit{preferential attachment} to an evolving network model and showed how this resulted in the degree distributions becoming scale-free for long enough times. For this to work, attachment had to be linear -- i.e., the probability a node with degree $k$ has of receiving a new edge is $\pi(k)\sim k+q$. This results in scale-free stationary degree distributions with an exponent $\gamma=3-q$.

Preferential attachment seems to be behind the emergence of many real-world,
continuously growing networks. However, not all networks in which some nodes at times gain (or loose) new edges have a continuously growing number of nodes. For example, a given group
of people may form an evolving social network \cite{Kossinets} in which the
edges represent friendship. Preferential attachment may be relevant here -- the more people you know, the more likely it is that you will be introduced
to someone new -- but probabilities are not expected to depend linearly on
degree. For instance, there may be saturations (highly connected people might
become less accessible), threshold effects (hermits may be prone to antisocial
tendencies), and other non-linearities. The brain may also be a relevant case.
Once formed, the number of neurons does not seem to continually augment, and
yet its structural topology is dynamic \cite{Klintsova}. Synaptic growth and
dendritic arborization have been shown to increase with electric stimulation
\cite{Lee,DeRoo} -- and, in general, the more connected a neuron is, the more
current it receives from the sum of its neighbours.

Barab\'{a}si and Albert showed that both (linear) preferential attachment and
an ever-growing number of nodes are needed for scaling to emerge in their
model. In a fixed population, their mechanism would result in a
fully-connected network. However, this is not normally observed in real
systems. Rather, just as some new edges sprout, others disappear -- less used
synapses suffer atrophy, unstimulating friendships wither. Often, the numbers
of both nodes and edges remain roughly constant. The same authors did
therefore extend their model so as to include the effects of
\textit{preferential rewiring} (which could be applied to fixed-size
networks), although again probabilities depended linearly on node degree
\cite{Albert}. Another mechanism which (roughly) maintains constant the
numbers of nodes and edges is node fusing \cite{Thurner}, once more according
to linear probabilities. As to nonlinear preferential attachment, the
(growing) BA model was extended to take power-law probabilities into account
\cite{Krapivsky}, although the solutions are only scale free for the linear case.

In this note we present an evolving network model with preferential rewiring
according to nonlinear (power-law) probabilities. The number of nodes and
edges is conserved but the topology evolves, arriving eventually at a
macroscopically (nonequilibrium) stationary state -- as described by global
properties such as the degree distribution. Depending on the exponents chosen
for the rewiring probabilities, the final state can be either fairly
homogeneous, with a typical size, or highly heterogeneous, with the emergence
of starlike structures. In the critical case marking the transition between
these two regimes, the degree distribution is shown to follow a nonlinear
diffusion equation. This describes a tendency towards stationary states that are
characterized either by scale-free or by mixed scale-free distributions,
depending on parameters.

Our model consists of a random network with $N$ nodes of respective degree
$k_{i},$ $i=1,2,...,N,$ and $%
\frac12 N\left\langle k\right\rangle $ edges. Initially, the degrees have a given
distribution $p(k,t=0)$. At each time step, one node is chosen with a
probability which is a function of its degree, $\rho(k_{i})$. One of its edges
is then chosen randomly and removed from it, to be reconnected to another node
$j$ chosen according to a probability $\pi(k_{j})$. That is, an edge is broken
and another one is created, and the total number of edges, as well as the
total number of nodes, is conserved. The functions $\pi(k)$ and $\rho(k)$ are
arbitrary, but we shall explicitly illustrate here $\pi(k_{i})\sim
k_{i}^{\alpha}$ and $\rho(k_{i})\sim k_{i}^{\beta}$ that capture the essence
of a wide class of nonlinear monotonous response functions and are easy to
handle analytically.

The probabilities $\pi$ and $\rho$ a given node has, at each time step, of
increasing or decreasing its degree can be interpreted as transition
probabilities between states. The expected value of the increment in a given
$p(k,t)$ at each time step, $\Delta p(k,t)$, may then be written as%

\begin{align}
\frac{\partial p(k,t)}{\partial t}  &  =(k-1)^{\alpha}\,\bar{k}_{\alpha}%
^{-1}p(k-1,t)\nonumber\\
&  +(k+1)^{\beta}\,\bar{k}_{\beta}^{-1}\,p(k+1,t)\label{master}\\
&  -\left(  k^{\alpha}\,\bar{k}_{\alpha}^{-1}+k^{\beta}\,\bar{k}_{\beta}%
^{-1}\right)  \,p(k,t),\nonumber
\end{align}
where $\bar{k}_{a}=\bar{k}_{a}\left(  t\right)  =\sum_{k}k^{a}p(k,t).$
If it exists, any stationary solution must satisfy the condition
$p_{\text{st}}(k+1)\,(k+1)^{\beta}\,\bar{k}_{\alpha}^{\text{st}}=p_{\text{st}%
}(k)\,k^{\alpha}\,\bar{k}_{\beta}^{\text{st}}$ which, for $k\gg 1$, implies
that
\begin{equation}
\frac{\partial p_{\text{st}}(k)}{\partial k}=\left(  \frac{\bar{k}_{\alpha
}^{\text{st}}}{\bar{k}_{\beta}^{\text{st}}}\frac{k^{\alpha}}{(k+1)^{\beta}%
}-1\right)  p_{\text{st}}(k). \label{det_bal}%
\end{equation}
Therefore, the distribution will have an extremum at 
$k_{e}=\left(  \bar
{k}_{\beta}^{\text{st}}/\bar{k}_{\alpha}^{\text{st}}\right)  ^{\frac{1}%
{\alpha-\beta}}$ (where we have approximated $k_{e}\simeq k_{e}+1$). If
$\alpha<\beta$, this will be a maximum, signalling the peak of the
distribution. On the other hand, if $\alpha>\beta$, $k_{e}$ will correspond to
a minimum. Therefore, most of the distribution will be broken in two parts,
one for $k<k_{e}$ and another for $k>k_{e}$. The critical case for
$\alpha=\beta$ will correspond to a monotonously decreasing stationary
distribution, but such that $\mbox{lim}_{k\rightarrow\infty}\partial
p_{st}(k)/\partial k=0$. In fact, Eq. (\ref{master}) is for this situation
($\alpha=\beta$) the discretised version of a nonlinear diffusion equation,
\begin{equation}
\frac{\partial p(k,\tau)}{\partial\tau}=\frac{\partial^{2}}{\partial k^{2}%
}[k^{\alpha}p(k,\tau)], \label{diff}%
\end{equation}
after dynamically modifying the time scale according to $\tau=t/\bar
{k}_{\alpha}\left(  t\right)  $. Ignoring, for the moment, border effects, the
solutions of this equation are of the form
\begin{equation}
p_{\text{st}}(k)\sim Ak^{-\alpha}+Bk^{-\alpha+1}, \label{sol}%
\end{equation}
with $A$ and $B$ constants. If $\alpha>2$, then given $A$ we can always find a
$B$ which allows $p_{\text{st}}(k)$ to be normalized in the thermodynamic
limit \cite{footnote}. For example, if the lower limit is $k\geq1$, then
$B=(\alpha-2)\left[  1-A/(\alpha-1)\right]  $. However, if $1<\alpha\leq2$,
then only $A$ can remain non-zero, and $p_{\text{st}}(k)$ will be a pure power
law. For $\alpha\leq1$, both constants must tend to zero as $N\rightarrow
\infty$.%
\begin{figure}
[tbh]
\begin{center}
\includegraphics[
width=7.95cm
]%
{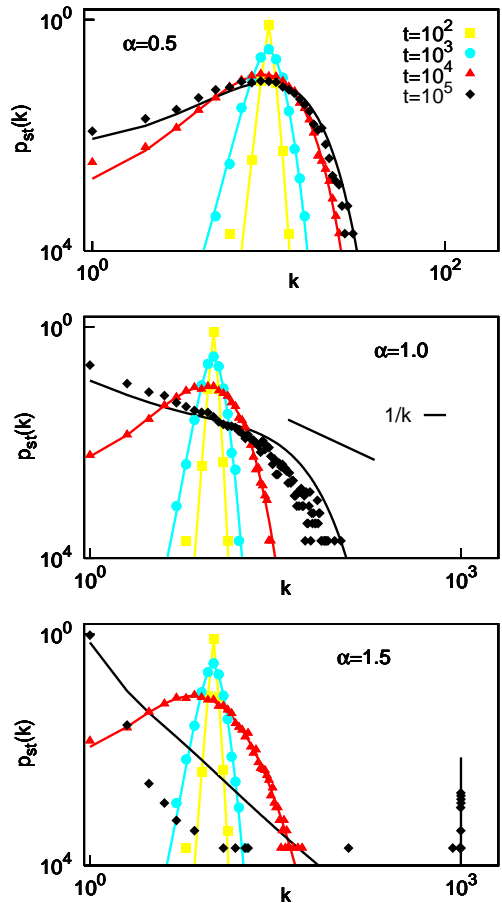}
\caption{(Color online) Degree distribution $p(k,t)$ at four different stages of evolution:
$t=10^{2}$ [(yellow) squares], $10^{3}$ [(blue) circles], $10^{4}$ [(red) triangles)] and $10^{5}$ MCS [(black) diamonds]. From top to bottom panels, subcritical ($\alpha=0.5$), critical ($\alpha=1$) and supercritical ($\alpha=1.5$) rewiring exponents. Symbols from MC simulations
and corresponding solid lines from numerical integration of Eq. (\ref{master}). $\beta=1$, $\langle k\rangle=10$ and $N=1000$ in all cases.}%
\label{fig1}%
\end{center}
\end{figure}
In finite networks, no node can have a degree larger than $N-1$ or lower than
$0$. In fact, one would usually wish to impose a minimum nonzero degree, e.g.
$k\geq1$. The temporal evolution of the degree distribution is illustrated in
Fig. \ref{fig1}. This shows the result of integrating Eq. (\ref{master}) for
$k\geq1,$ different times, $\beta=1,$ and three different values of $\alpha$,
along with the respective values obtained from Monte Carlo simulations.%

\begin{figure}
[tbh]
\begin{center}
\includegraphics[
width=9.cm
]%
{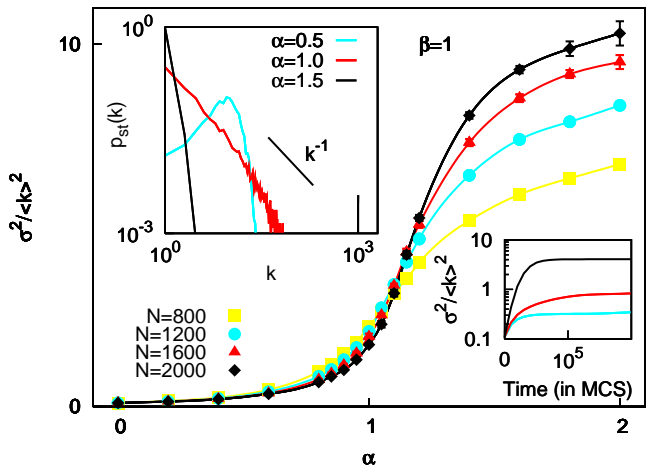}
\caption{(Color online) Adjusted variance $\sigma^{2}/\langle k\rangle^{2}$ of the degree
distribution after $2\times10^{5}$ MCS against $\alpha$, as obtained from
MC simulations, for system sizes $N=800$ [(yellow) squares], $1200$ [(blue) circles], $1600$ [(red) triangles] and $2000$ [(black) diamonds]. Top left inset shows final degree distributions for $\alpha=0.5$ [light gray (blue)], $1$ [dark gray (red)] and $1.5$ (black), with $N=1000$. Bottom right inset shows typical time series of $\sigma^{2}/\langle k\rangle^{2}$ for the same three values of $\alpha$ and $N=1200$. In all cases, $\beta=1$ and $\langle k\rangle=10$.}%
\label{fig2}%
\end{center}
\end{figure}

The main result may be summarized as follows. For $\alpha<\beta$, the network
will evolve to have a characteristic size, centred around $\langle k\rangle$.
At the critical case $\alpha=\beta$, all sizes appear, according either to a
pure or a composite power law, as detailed above.

If we impose, say, $k\geq1$, then starlike structures will emerge, with a great many nodes connected to just a few
hubs \footnote{There is a finite-size effect not taken into account by the theory -- but relevant when $\alpha>\beta$ -- which provides a natural lower cutoff for $p_{st}(k)$: if there are, say, $m$ nodes which are connected to the whole network, then the minimum degree a node can have is $m$.}.

Figure \ref{fig2} illustrates the second order phase transition undergone by
the variance of the final (stationary) degree distribution, depending on the
exponent $\alpha$, where $\beta$ is set to unity. It should be mentioned that
this particular case, $\beta=1$, corresponds to edges being chosen at random
for disconnection, since the probability of a random edge belonging to node
$i$ is proportional to $k_{i}$.

This topological phase transition is similar to the ones that have been described in equilibrium network ensembles defined via an energy function, in the so-called \textit{synchronic} approach to network analysis \cite{Farkas,Park,Burda,Derenyi}. However, our (nonequilibrium) model does not come within the scope of this body of work, since the rewiring rates cannot, in general, be derived from a potential. Furthermore, we are here concerned with the time evolution rather than the stationary states, making our approach \textit{diachronic.} 

Summing up, in spite of its simplicity, our model captures the essence of many
real-world networks which evolve while leaving the total numbers of nodes and
edges roughly constant. The grade of heterogeneity of the stationary
distribution obtained is seen to depend crucially on the relation between the
exponents modelling the probabilities a node has of obtaining or loosing a new
edge. It is worth mentioning that the heterogeneity of the degree distribution
of a random network has been found to determine many relevant behaviours and
magnitudes such as its clustering coefficient and mean minimum path
\cite{Newman_rev}, critical values related to the dynamics of excitable
networks \cite{Johnson}, or the synchronisability for systems of coupled
oscillators (since this depends on the spectral gap of the Laplacian matrix)
\cite{Barahona}.

The above shows how scale-free distributions, with a range of
exponents, may emerge for nonlinear rewiring, although only in the critical
situation in which the probabilities of gaining or loosing edges are the same.
We believe that this non-trivial relation between the microscopic rewiring
actions (governed in our case by parameters $\alpha$ and $\beta$) and the emergent
macroscopic degree distributions could shed light on a class of biological,
social and communications networks.

This work was supported by Junta de Anadaluc\'{i}a project FQM-01505 and by Spanish MEC-FEDER
project FIS2009-08451


\begin{thebibliography}{99}                                                                                               %


\bibitem {gen}S. Boccaletti \textit{et al.}, \textsl{Phys. Rep.}
\textbf{424}, 175 (2006)

\bibitem {gen1}{A. Arenas \textit{et al., } \textsl{Phys. Rep.} }\textbf{469}{, 93
(2008)}

\bibitem {gen2}J. Marro, J.J. Torres and J.M. Cortes, \textsl{J. Stat. Mech.: Theory and Experiment}, P02017 (2008)

\bibitem {Barabasi}A.-L. Barab\'{a}si and R. Albert, \textsl{Science
~\textbf{286}} 509--512 (1999)

\bibitem {Kossinets}G. Kossinets and D.J. Watts, \textsl{Science
~\textbf{311}}, 88--90 (2006)

\bibitem {Klintsova}A.Y. Klintsova and W.T. Greenough, \textsl{Current
Opinion in Neurobiology ~\textbf{9}}, 203--208 (1999)

\bibitem {Lee}K.S. Lee, F. Schottler, M. Oliver, and G. Lynch, \textsl{J. Neurophysiol.~\textbf{44}}, 247--258 (1980)

\bibitem {DeRoo}M. De Roo, P. Klauser, P. Mendez, L. Poglia, and D. Muller,
\textsl{Cerebral Cortex ~\textbf{18}} 151--161 (2008)

\bibitem {Albert}R. Albert and A.-L. Barab\'{a}si, \textsl{Phys. Rev. Lett. ~\textbf{85}}, 5234 (2000)

\bibitem {Thurner}S. Thurner, F. Kyriakopoulos, and C. Tsallis, 
\textsl{Phys. Rev. E. ~\textbf{76}}, 036111 (2007)

\bibitem {Krapivsky}P.L. Krapivsky, S. Redner, and F. Leyvraz,
\textsl{Phys. Rev. Lett. ~\textbf{85}}, 4629 (2000)

\bibitem {footnote}Although all moments of $k$ will diverge unless $B=0$.

\bibitem {Bianconi}G. Bianconi and A.-L. Barab\'{a}si, \textsl{Phys.
Rev. Lett. ~\textbf{86}}, 5632 (2001)

\bibitem{Farkas}I. Farkas, I. Der\'{e}nyi, G. Palla, and T. Vicsek,
Lect. Notes in Phys. ~\textbf{650}, 163 (2004)

\bibitem{Park}J. Park and M.E.J. Newman,
\textsl{Phys. Rev. E. ~\textbf{70}}, 066117 (2004)

\bibitem{Burda}Z. Burda, J. Jurkiewicz, and A. Krzywicki,
\textsl{Physica A ~\textbf{344}}, 56 (2004)

\bibitem{Derenyi} I. Der\'{e}nyi, I. Farkas, G. Palla, T. Vicsek,
\textsl{Physica A ~\textbf{344}}, 583 (2004)

\bibitem {Newman_rev}M.E.J. Newman, \textit{SIAM Reviews
}\textbf{45}, 167 (2003)

\bibitem {Johnson}S. Johnson, J. Marro, and J.J. Torres, \textsl{Europhys. Lett.} \textbf{83}, 46006 (2008)

\bibitem {Barahona}M. Barahona and L.M. Pecora, \textsl{Phys. Rev.
Lett. ~\textbf{89}}, 054101 (2002)


\end{thebibliography}
\end{document}